# Feasibility of a General-Purpose Deep Learning Dose Engine: A Multi-Site Validation Study

Yao Zhao, Ka Ho Tam, Raphael Douglas, Kyuhak Oh, Xin Wang, Ergys Subashi, Jinzhong Yang, Laurence Court, Dong Joo Rhee

Department of Radiation Physics, The University of Texas MD Anderson Cancer Center, Houston, TX 77030, USA

**Abstract**

**Background:** Conventional radiotherapy dose calculation algorithms exhibit significant variability in computational speed depending on anatomical complexity, creating a potential bottleneck for time-sensitive applications like online adaptive radiotherapy (ART). Furthermore, their non-differentiable nature limits the development of fully end-to-end automatic planning frameworks. Deep learning offers a solution by providing consistent inference performance and a differentiable framework essential for both rapid adaptation and optimization.

**Purpose**: The primary objective of this study was to develop a generalized deep learning-based dose calculation engine capable of accurate, site-independent dose prediction. By utilizing a beamlet-based input strategy, we aimed to establish a computationally consistent and differentiable dose module that enables end-to-end training for autoplanning, maintaining dosimetric accuracy across diverse anatomical geometries.

**Methods**: A dataset of 3,600 step-and-shoot IMRT and 3D-CRT plans (6 MV) was generated from 120 patients evenly distributed across six anatomical sites. We investigated two 3D convolutional neural network architectures—a standard U-Net and a coarse-to-fine Cascade U-Net—to predict 3D dose distributions directly from patient CT images and divergent MLC and jaw segment projections ("beamlets"). Models were trained using both Mean Squared Error (MSE) and Mean Absolute Error (MAE) loss functions. Performance was validated using 3D gamma analysis on an independent external cohort of 60 VMAT plans.

**Results:** The deep learning framework demonstrated high dosimetric accuracy across all tested configurations. The optimal model (U-Net trained with MAE loss) achieved a mean gamma passing rate of 98.9 ± 1.6% (3%/2mm, 10% threshold) on the independent test set, with the Cascade U-Net achieving a similarly high passing rate of 98.8 ± 1.6%. The model maintained robust performance across all six anatomical sites, with passing rates consistently exceeding 98%, demonstrating that the beamlet-based input strategy effectively generalizes to complex geometries without site-specific training.

**Conclusions:** We demonstrated that a single, site-independent deep learning model can calculate radiotherapy dose distributions with clinical accuracy. By effectively learning the relationship between patient anatomy, beam geometry, and dose distribution, this approach provides a computationally consistent and differentiable engine. This makes it highly suitable for integration into end-to-end automatic planning, as well as online ART and secondary dose verification workflows.

**Introduction**

Accurate dose calculation is essential for modern radiation therapy. In the context of highly modulated delivery techniques—specifically Intensity-Modulated Radiotherapy (IMRT) and Volumetric Modulated Arc Therapy (VMAT)—calculation algorithms are required to rigorously model radiation transport.[1,2] Currently, Monte Carlo (MC) methods serve as the reference standard due to their ability to explicitly model particle transport in heterogeneous media.[3] However, despite advances in GPU-based acceleration, the stochastic nature of MC necessitates simulating large numbers of particle histories to minimize statistical uncertainty, creating a substantial computational burden that limits its utility in time-critical workflows.[3–5] Conversely, analytical algorithms, such as Collapsed Cone Convolution (CCC),[6] and deterministic solvers are widely utilized in commercial Treatment Planning Systems (TPS) for their efficiency.[7,8] Yet, these methods often rely on approximations of lateral electron transport that can lead to dosimetric discrepancies, particularly at tissue interfaces with significant density gradients—such as the lung-tumor interface—or in the presence of strong magnetic fields used in MR-guided radiotherapy.[6,9,10] This trade-off between dosimetric accuracy and computational speed represents a critical bottleneck for the implementation of online Adaptive Radiotherapy (ART). As ART requires the re-optimization and recalculation of dose distributions on daily anatomy while the patient remains on the treatment couch, the latency associated with dose calculation directly impacts clinical throughput and increases the risk of intrafraction motion.[11–14] To address this challenge, deep learning has emerged as a promising approach for accelerating dose calculation by learning a direct mapping from beam-anatomy information to the three-dimensional dose distribution.[15,16]

Early work by Fan *et al.* proposed a data-driven dose calculation algorithm based on a deep U-Net that takes patient CT images and fluence-related inputs and outputs the corresponding dose distribution.[16] Their results showed that a trained network could reproduce TPS dose with sub-3% voxel errors for multiple treatment sites, suggesting that a deep learning model could serve as a surrogate dose engine. Subsequent studies extended this concept to more complex techniques and systems. For example, Kontaxis *et al.* introduced DeepDose, a deep learning dose engine trained on segment-wise MC dose to enable fast dose computation for prostate IMRT.[17] Related segment-/aperture-driven work has also been demonstrated for abdominal targets in a 1.5 T MR-guided setting.[18] Similarly, Xiao *et al.* demonstrated a synthetic-CT-free approach for MR-guided radiotherapy, directly mapping MR intensity to dose and thereby eliminating the errors associated with explicit electron density generation.[19] Beyond direct dose calculation, hybrid "boosting" strategies proposed by Xing *et al.* have successfully utilized hierarchically dense U-Nets to convert low-accuracy analog doses (e.g., AAA) into high-accuracy equivalents (e.g., Acuros XB), bridging the gap between speed and precision.[20] More recently, Liang *et al.* developed a deep learning–based dose calculation method for VMAT, using a 3D U-Net to calculate dose from projected fluence maps, CT, and radiological depth information.[21] The model achieved good agreement with a clinical dose engine and was proposed as a fast surrogate during VMAT optimization.

However, despite these promising advancements, significant challenges remain in the clinical translation of deep learning-based dose engines. First, the majority of existing models are trained on narrow, site-specific datasets such as exclusively abdominal targets on an MR-linac or single-site VMAT cases.[18,19] Therefore, the feasibility of a single deep learning model capable of maintaining dosimetric accuracy across diverse anatomical geometries and varying tissue heterogeneities remains largely unexplored. In addition, many existing approaches are constrained by dependencies on specific delivery techniques or systems. For instance, fluence-based models rely on plan-level derived quantities—such as total projected fluence or radiological depth.[16] While these are effective for specific tasks, these inputs implicitly encode machine information, limiting their application to more general applications. Similarly, hybrid "boosting" strategies necessitate an initial low-accuracy dose calculation. This requirement creates a dependency on commercial TPS, preventing its use as a standalone dose-calculation engine.[20] Finally, although segment-based approaches have successfully moved beyond fluence maps by utilizing 3D aperture inputs, they typically use single-stage U-Net architectures.[21] A primary limitation of such single-stage networks is the inherent difficulty in simultaneously capturing global anatomical information and local details. In steep dose gradient regions—such as the penumbra or near small organs-at-risk—these networks often yield "blurred" calculations.

In this work, we address these limitations by developing a beamlet-based, 3D convolutional deep learning dose engine that calculates dose directly from patient CT and divergent MLC and jaw projections. Unlike previous methods, our method does not rely on TPS-specific intermediate dose calculations or segment-wise Monte Carlo pre-calculations. By training on a large, heterogeneous dataset VMAT and 3D-CRT plans spanning six anatomical sites, our model is explicitly designed to learn a generalizable mapping between patient anatomy, aperture geometry, and dose. As a result, the model is not restricted to a single disease site, delivery technique, or vendor platform. We evaluated the models using 3D gamma analysis on an independent cohort of VMAT plans. Crucially, the proposed deep learning framework is inherently differentiable, enabling the direct backpropagation of gradients from dosimetric loss functions to machine parameters. This capability positions the module as a foundational component for end-to-end automatic planning training, while also providing a computationally consistent and independent dose calculation tool suitable for integration into open-source planning systems and adaptive radiotherapy workflows.

**Methods**

**Patient Cohort and Data Generation**

The dose calculation model was trained using a retrospective cohort of 120 patients treated at MD Anderson Cancer Center, balanced evenly across six anatomical sites: brain, head-and-neck, breast, thoracic, gastrointestinal, and genitourinary. For each patient, an in-house autoplanning script in RayStation was utilized to generate 20 step-and-shoot IMRT plans and 10 3D-CRT plans using 6 MV photon beams.[22,23] To ensure data diversity, plans included randomized isocenter placement within the target, variable gantry angles, and randomized dose constraints (for IMRT). Individual beamlets were extracted from these plans to serve as model inputs, yielding a total dataset of 201,214 beamlets. The data were partitioned at the patient level into training, validation, and testing sets (8:1:1 ratio).

**Data Preprocessing**

As shown in Figure 1, each input sample consisted of four 3D matrices with dimensions of 192 x 192 x 192 and an isotropic voxel size of 3 mm, centered at the machine isocenter: the simulation CT, the divergent MLC projection, the divergent Jaw projection, and the ground-truth 3D dose distribution per unit MU. CT numbers were converted to electron density using the scanner-specific calibration table; values outside the external body contour were masked to zero density. To match the clinical dose calculation geometry, a digital couch structure was inserted into the CT image using the standard physical density values from the TPS. Furthermore, any manual density overrides present in the original clinical plans were replicated in the input CT images. The MLC and Jaw projections utilized normalized voxel values ranging from 0 (fully shielded) to 1 (fully open). For voxels partially traversed by the collimator leaves, values were calculated proportional to the geometric aperture opening using a raytracing technique.

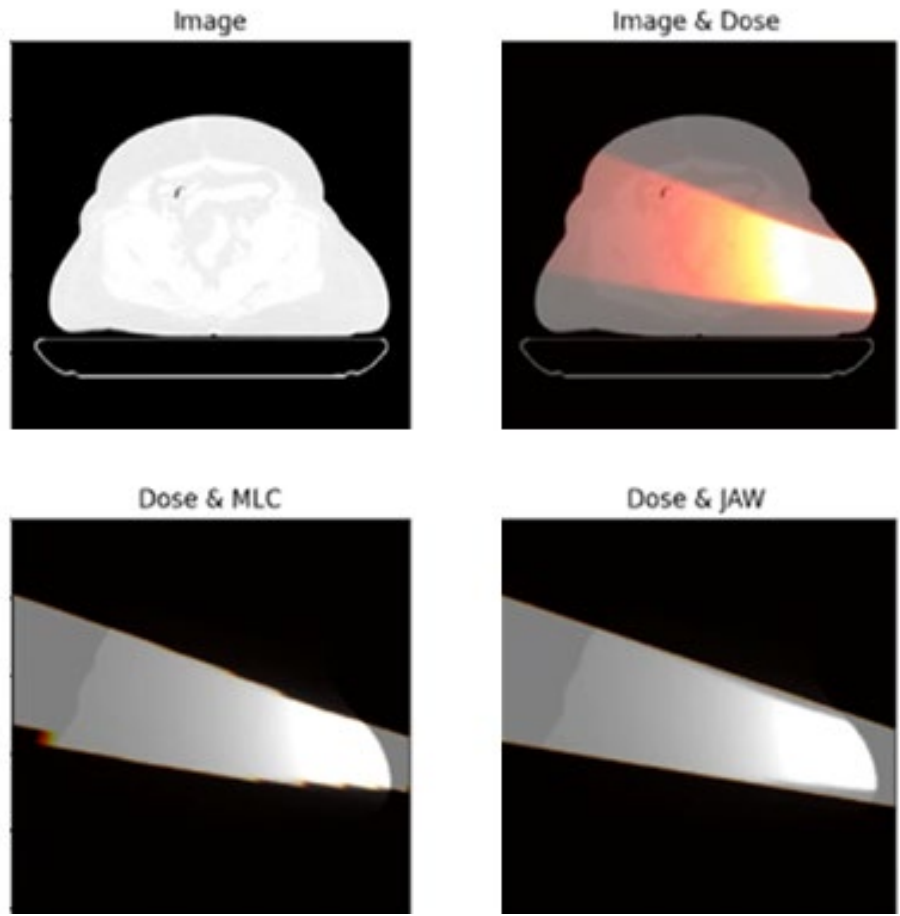

**Figure 1.** Representative transverse image of model inputs. The figure illustrates the spatial co-registration of the input channels for a single beamlet. The top row shows the planning CT, both alone (left) and overlaid with the ground-truth dose distribution (right). The bottom row displays the ray-traced divergent projections for the MLC (left) and Jaws (right), which define the geometric beam aperture; both are overlaid with the dose distribution. All data are derived from 192 x 192 x 192 volumetric matrices with a 3 mm isotropic resolution.

**Deep Learning Architecture**

We investigated and compared two convolutional neural network-based deep learning architectures for dose prediction: a baseline 3D U-Net[24] and a 3D Cascade U-Net[25]. The fundamental building block for both networks was a 3D Convolutional Block consisting of two 3×3×3 convolutional layers. To handle the small batch sizes typical of 3D medical imaging, we utilized Instance Normalization with learnable affine parameters, followed by a LeakyReLU[26] activation function. Downsampling in the encoding path was performed using strided convolutions (kernel size 2, stride 2), while the decoding path utilized transposed convolutions for upsampling.

The 3D Cascade U-Net employed a "coarse-to-fine" strategy consisting of two nested U-Net-like structures. The first network, serving as the coarse stage, featured an encoder depth of 3 levels with a base filter count of 16. The second network, serving as the fine stage, was deeper, with an encoder depth of 4 levels and a base filter count of 32. Unlike standard cascades that only forward the predicted output probability map, our architecture propagated the dense feature map (16 channels) from the last decoder layer of the coarse stage. These features were concatenated with the original 3-channel input images to form a 19-channel input tensor for the fine stage. This design allowed the fine network to leverage rich contextual information learned by the coarse network to correct residuals and refine high-gradient dose regions.

**Implementation and Evaluation**

The models were implemented using the PyTorch framework and trained on a multi-GPU cluster. We utilized a Distributed Data Parallel (DDP) strategy to maximize computational efficiency. Optimization was performed using the Adam optimizer with a fixed learning rate of

$1\times10^{-4}$. To optimize the Cascade network effectively, we employed a deep supervision strategy where the total loss function was defined as a weighted sum of the coarse and fine stage losses (L_total = 0.5×L_coarse + L_fine). This supervision ensured that the coarse network learned meaningful representations early in the training process. The models were trained for 100 epochs, and the model state yielding the lowest validation loss was saved as the optimal checkpoint for testing.

To quantitatively assess the agreement between the deep learning-predicted dose and the ground-truth distributions (defined as the dose calculated by the clinical TPS), we employed 3D Gamma index analysis[27]. A global gamma evaluation was performed using four specific criteria: 3%/2mm, 3%/1mm, 2%/2mm, 2%/1mm. To focus on clinically relevant volumes, a lower dose threshold of 10% of the maximum dose was applied and passing rates were calculated as the percentage of voxels within the body contour satisfying the condition $\gamma < 1$. Following internal validation, we applied these metrics to an independent test dataset to evaluate model generalizability. This external cohort consisted of 60 clinical plans from both MD Anderson and the Radiation Planning Assistant[28], a web-based automated planning platform. The dataset consisted of 60 VMAT plans covering head-and-neck, prostate, gynecologic, thoracic, breast, and gastrointestinal cancer sites, evenly distributed with 10 patients per site.

**Results**

Table 1 summarizes the gamma passing rates (GPR) for the entire independent test cohort (n = 60) across different model architectures and loss functions. Under the standard clinical criteria of 3%/2mm, the mean GPRs were 98.8%, 98.8%, 98.9%, and 98.3% for the Cascade U-Net (MAE), Cascade U-Net (MSE), U-Net (MAE), and U-Net (MSE), respectively. As illustrated in the boxplots in Figure 2, models trained with the MAE loss function consistently exhibited marginally superior performance compared to those trained with MSE, particularly in stricter criteria 2%/1mm. When comparing architectures using the optimal MAE loss, no significant difference was observed between the Cascade U-Net and the baseline U-Net, with both achieving clinical acceptance levels (>95%) in the vast majority of cases.

The dosimetric accuracy stratified by anatomical site is detailed in Table 2 and Figure 3. The deep learning models demonstrated the highest agreement for pelvic sites, achieving mean GPRs 3%/2mm) of >99% for Cervical and Prostate cases. Performance remained robust for Head-and-Neck and Thoracic plans, with passing rates exceeding 98.6%. While slightly lower agreement was observed for Breast and Gastrointestinal sites, the models maintained high clinical reliability. Notably, even for the most challenging sites, the minimum GPR for the best-performing model remained above 93%.

Table 1. Overall Gamma Analysis. Summary of gamma passing rates (mean ± standard deviation) for the independent test dataset (n=60). Comparison across four gamma criteria, two deep learning architectures (Cascade U-Net vs. U-Net), and two loss functions (MAE vs. MSE).

| Model & Loss | 3%/2mm (mean ± stdev) | 3%/1mm (mean ± stdev) | 2%/2mm (mean ± stdev) | 2%/1mm (mean ± stdev) |
|---|---|---|---|---|
| Cascade U-Net with MAE | 98.8 ± 1.6 | 96.6 ± 3.2 | 97.0 ± 3.1 | 91.7 ± 6.3 |
| Cascade U-Net with MSE | 98.8 ± 1.5 | 96.3 ± 4.1 | 97.0 ± 3.2 | 91.3 ± 7.7 |
| UNet with MAE | 98.9 ± 1.6 | 96.6 ± 3.2 | 97.1 ± 3.2 | 91.6 ± 6.5 |
| UNet with MSE | 98.3 ± 2.1 | 95.7 ± 3.9 | 95.7 ± 4.2 | 89.2 ± 7.9 |

Table 2. Site-Specific Gamma Analysis. Gamma passing rates stratified by anatomical site. Results are reported as mean ± standard deviation for the test dataset (n=10 per site).

| Sites | Model & Loss | 3%/2mm (mean ± stdev) | 3%/1mm (mean ± stdev) | 2%/2mm (mean ± stdev) | 2%/1mm (mean ± stdev) |
|---|---|---|---|---|---|
| Breast | Cascade U-Net with MAE | 97.5 ± 1.6 | 94.1 ± 2.4 | 93.7 ± 2.9 | 85.7 ± 4.1 |
| | Cascade U-Net with MSE | 97.6 ± 1.8 | 93.1 ± 4.7 | 94.1 ± 3.0 | 84.9 ± 6.5 |
| | UNet with MAE | 98.2 ± 1.3 | 95.3 ± 2.0 | 95.3 ± 3.0 | 88.0 ± 4.0 |
| | UNet with MSE | 96.5 ± 2.3 | 92.2 ± 4.0 | 91.8 ± 4.5 | 82.0 ± 7.6 |
| Gastrointestinal | Cascade U-Net with MAE | 98.2 ± 2.1 | 96.2 ± 3.4 | 96.1 ± 3.5 | 91.4 ± 6.3 |
| | Cascade U-Net with MSE | 98.5 ± 1.6 | 96.4 ± 2.4 | 96.4 ± 2.8 | 91.4 ± 5.7 |
| | UNet with MAE | 98.0 ± 2.7 | 95.9 ± 4.2 | 95.9 ± 4.5 | 91.3 ± 7.6 |
| | UNet with MSE | 97.9 ± 2.3 | 95.6 ± 3.8 | 95.1 ± 4.5 | 89.5 ± 7.9 |
| Cervical | Cascade U-Net with MAE | 99.8 ± 0.2 | 98.9 ± 0.7 | 99.4 ± 0.6 | 96.9 ± 2.0 |
| | Cascade U-Net with MSE | 99.6 ± 0.3 | 98.7 ± 1.0 | 98.7 ± 1.3 | 95.6 ± 3.7 |
| | UNet with MAE | 99.2 ± 1.8 | 97.5 ± 3.8 | 97.7 ± 3.8 | 92.9 ± 8.5 |
| | UNet with MSE | 99.3 ± 0.9 | 98.0 ± 1.7 | 97.6 ± 2.1 | 93.1 ± 4.8 |
| Head-and-neck | Cascade U-Net with MAE | 99.3 ± 0.6 | 95.4 ± 2.1 | 97.8 ± 1.3 | 89.3 ± 4.1 |
| | Cascade U-Net with MSE | 99.2 ± 0.6 | 95.2 ± 2.4 | 97.5 ± 1.6 | 89.3 ± 4.8 |
| | UNet with MAE | 99.0 ± 1.2 | 95.2 ± 2.6 | 97.2 ± 2.3 | 89.1 ± 4.5 |
| | UNet with MSE | 98.7 ± 1.8 | 94.3 ± 3.3 | 96.6 ± 2.7 | 87.3 ± 4.9 |
| Prostate | Cascade U-Net with MAE | 99.4 ± 1.4 | 97.8 ± 4.6 | 98.6 ± 2.7 | 95.0 ± 8.1 |
| | Cascade U-Net with MSE | 99.1 ± 2.2 | 97.4 ± 6.3 | 98.1 ± 4.6 | 94.6 ± 11.5 |
| | UNet with MAE | 99.6 ± 0.8 | 98.6 ± 2.7 | 98.8 ± 2.2 | 96.3 ± 6.0 |
| | UNet with MSE | 99.4 ± 1.3 | 98.3 ± 3.2 | 98.3 ± 3.2 | 95.3 ± 7.0 |
| Thoracic | Cascade U-Net with MAE | 98.6 ± 2.0 | 97.1 ± 2.6 | 96.5 ± 3.5 | 91.8 ± 5.6 |
| | Cascade U-Net with MSE | 98.8 ± 1.4 | 96.8 ± 3.6 | 96.9 ± 3.2 | 91.9 ± 8.0 |
| | UNet with MAE | 99.3 ± 0.6 | 97.3 ± 2.1 | 97.5 ± 2.1 | 92.3 ± 5.6 |
| | UNet with MSE | 97.9 ± 2.4 | 95.6 ± 4.3 | 94.6 ± 4.4 | 88.2 ± 8.4 |

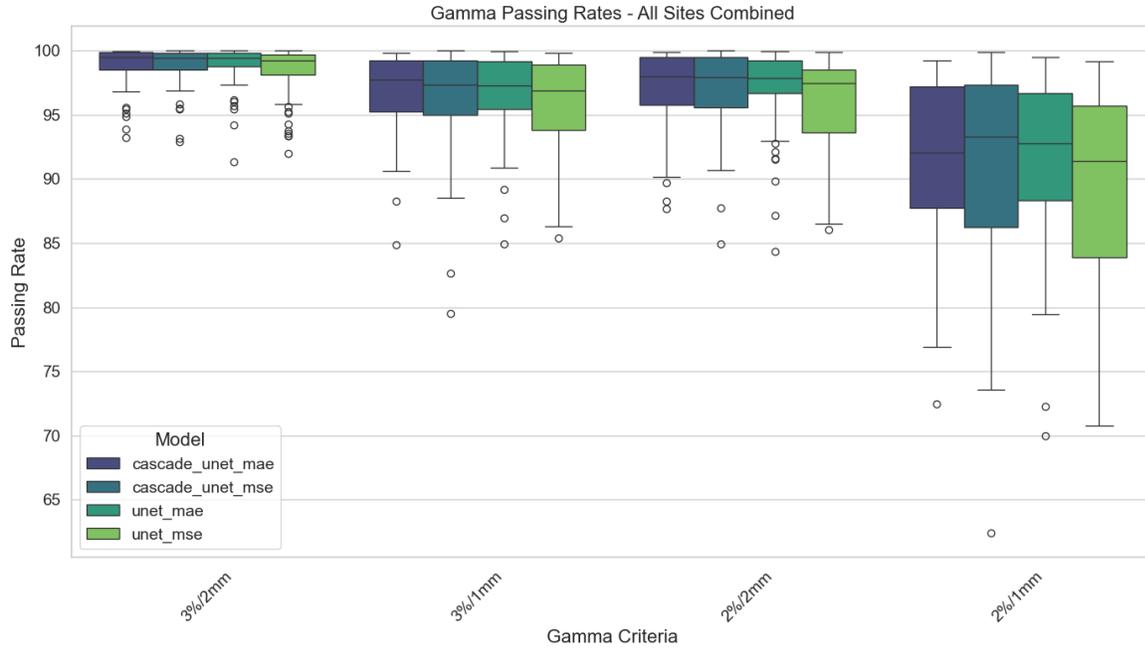

**Figure 2.** Boxplots illustrating the dosimetric accuracy of the four tested model configurations across varying gamma criteria (3%/2mm, 3%/1mm, 2%/2mm, and 2%/1mm).

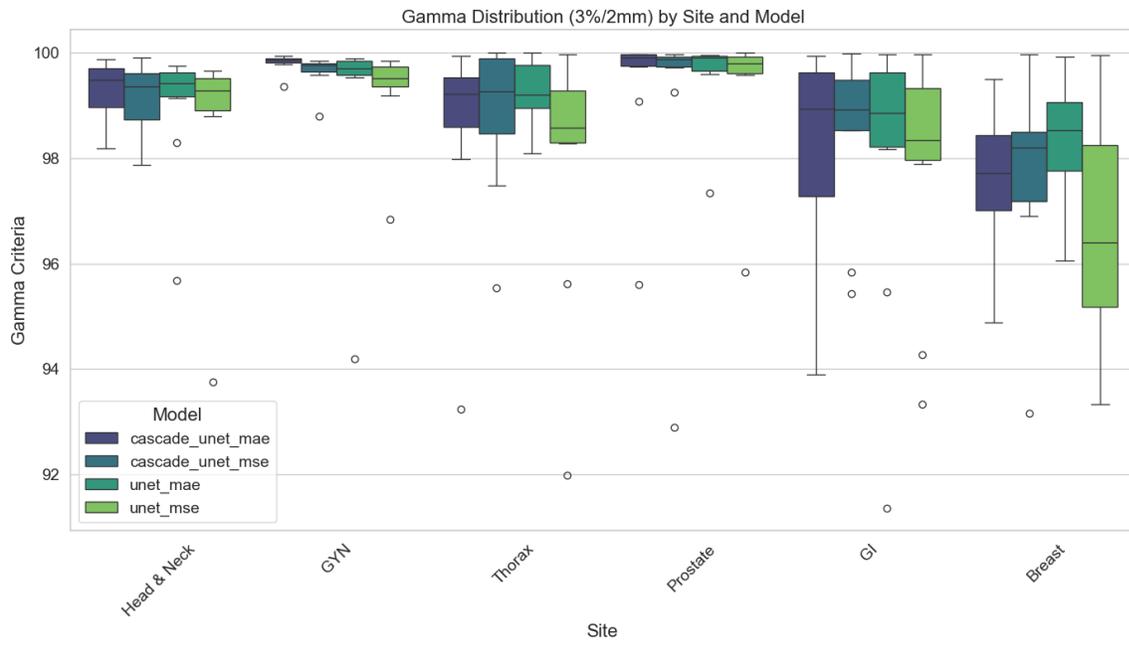

**Figure 3**. Boxplots showing the gamma passing rates (3%/2mm criteria) stratified by the six anatomical cancer sites.

To better understand the model's limitations, we performed a qualitative analysis of the 'worst-case' outliers. Figure 4 illustrates the dose distributions and difference maps for the cases with the lowest gamma passing rates. In breast cancer plans, discrepancies were primarily localized to the buildup region near the skin surface and the interfaces surrounding high-density surgical clips. These errors likely stem from the steep dose gradients and scattering effects associated with abrupt density changes. Similarly, in thoracic cases, prediction accuracy was challenged by the significant tissue heterogeneity at the soft tissue-lung interface.

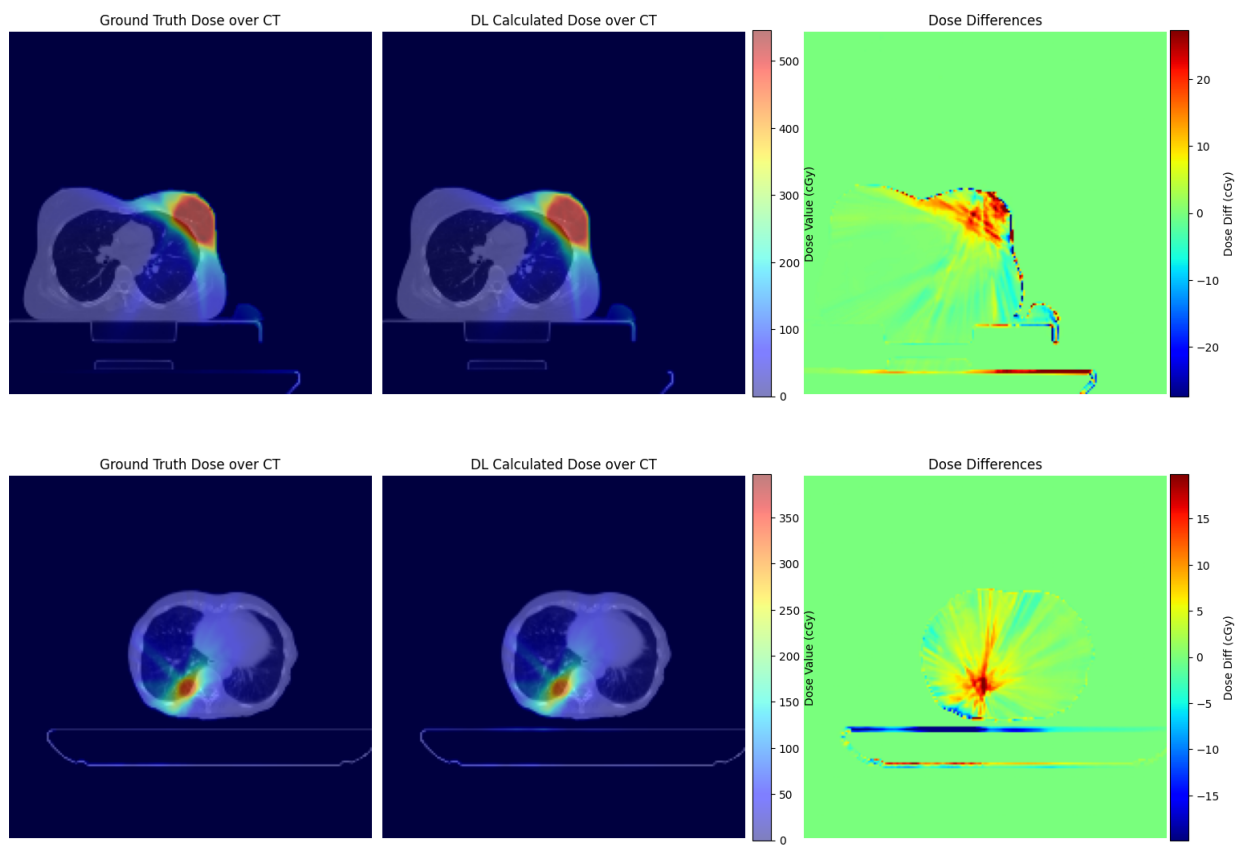

**Figure 4**. Visualization of worst-case scenarios. Comparison of the ground-truth dose (left), deep learning-predicted dose (center), and pixel-wise dose difference (right) for the lowest-performing cases in the test set. The rows correspond to: (Top) Breast (3%/2mm GPR: 94.9%) and (Bottom) Thoracic (3%/2mm GPR:

93.2%). Note the localized errors near the skin surface for the breast case and at the tumor-lung interface for the thoracic case.

**Discussion**

In this study, we successfully developed and evaluated a beamlet-based deep learning dose calculation engine that calculates 3D dose directly from patient CT and divergent MLC/jaw aperture projections. The model was trained on a diverse dataset of 3D-CRT and IMRT treatment plans spanning six anatomical sites, making it effectively site-independent. The deep learning-based dose calculations closely matched the clinical TPS calculations, achieving very high gamma index passing rates – on average above 98% for a 3%/2 mm criterion, and remaining high even under stricter criteria. This indicates that the deep learning engine can reproduce dose calculations with near TPS-level accuracy. The model performed robustly across all tested sites, with the best accuracy observed in pelvic cases, likely due to the relatively uniform anatomy in pelvic regions. Minor localized dose discrepancies were observed in challenging scenarios such as near air–tissue interfaces (e.g. skin surface, lung-tissue interface) and around high-density surgical clips, where the network occasionally underperformed due to steep density gradients. Overall, the key finding is that a single deep learning model can accurately calculate 3D dose distributions for multiple treatment sites and beam arrangements, demonstrating feasibility as a general-purpose dose calculation engine.

A major strength of this work is the generality and flexibility of the proposed dose engine. Training on multiple tumor sites (brain, head-and-neck, thorax, abdomen, pelvis, etc.) with both 3D-CRT and IMRT techniques allowed the model to learn broadly applicable dose deposition features. This site-independent training strategy means the same model can be applied to new patients across different disease sites without needing site-specific fine-tuning, an advantage over typical models limited to one region. Additionally, the use of divergent aperture projections ("beamlets") as inputs makes the network beam-geometry independent – it can handle arbitrary beam arrangements or number of fields by summing contributions from individual beamlets. Crucially, by consistently aligning the machine isocenter to the geometric center of the input images and apertures, the deep learning model can effectively internalize fixed physical parameters such as the beam profile and Source-to-Axis Distance (SAD), simplifying the learning task. This approach allows the 3D U-Net architecture to effectively capture physical dose spread characteristics, such as inverse-square fall-off and tissue attenuation, directly from the data. Another strength is the accuracy and reliability of the dose calculation. While we hypothesized that a Cascade architecture might be necessary for fine detail, our results demonstrated that the standard 3D U-Net was sufficient to achieve high gamma pass rates, even under strict criteria. This indicates that the high-resolution divergent input data, rather than architectural complexity, is the primary driver of accuracy. The average >98% gamma passing at 3%/2 mm indicates the deep learning-based calculations are nearly indistinguishable from conventional calculations for most voxels. Such accuracy was consistent across test cases, highlighting reliability.

Our results are consistent with the accuracy reported by prior deep learning dose calculation studies. Early works established that 3D U-Net models could predict radiotherapy dose with high accuracy; however, these models were predominantly site-specific or plan-specific. They are typically restricted to a single anatomical region (e.g., head-and-neck) or fixed beam arrangements. In contrast, our model distinguishes itself through its site-independent design. By leveraging divergent beamlet projections, the network learns fundamental physics-based dose deposition features that generalize across diverse anatomies and beam configurations.

A comparable concept was recently explored by Rousselot *et al.*, who used beamlet decomposition to train a generic dose engine adaptable to various X-ray beam geometries. They similarly found that a single model could handle different beam orientations and even imaging modalities without retraining. Our work reinforces this by demonstrating high accuracy across six anatomical sites with one single model. In terms of quantitative performance, the >98% gamma passing rate (3%/2mm) achieved in this study compares favorably to literature benchmarks. For example, some studies reported ~95–99% gamma passing at 3%/3mm for deep-learned dose calculations, and even ~98% at 1%/1mm in specialized conversion tasks. Our results under tighter 3%/2mm criteria underscore the accuracy of the beamlet-based approach. Our findings suggest that a physics-informed, beamlet input strategy can achieve state-of-the-art accuracy while offering greater generalizability than many earlier CNN-based methods.

In conclusion, this work demonstrates that a beamlet-based deep learning approach effectively overcomes the traditional trade-off between computational speed and dosimetric accuracy. By utilizing divergent projection inputs, the developed model achieves robust site-independence, demonstrating the ability to generalize across heterogeneous anatomies without site-specific fine-tuning. However, the study identified specific limitations. Qualitative analysis revealed minor localized discrepancies in regions characterized by sharp density gradients. These findings suggest that while the global dose distribution is accurate, further methodological development is required to fully resolve dose perturbation at substantial heterogeneity boundaries. Furthermore, the current model was trained and validated exclusively using 6 MV photon beams. The applicability of this approach to other beam energies commonly utilized in clinical practice, such as 15 MV or Flattening Filter Free (FFF) modes (e.g., 6FFF, 10FFF), remains to be investigated. Future efforts will focus on extending the model's capabilities to these diverse energy spectra and addressing heterogeneity challenges. Ultimately, the computational efficiency and generalizability of this approach position it as a promising engine for open-source treatment planning and time-critical online adaptive radiotherapy applications

**Conclusion**

We have validated a beamlet-based deep learning dose engine that achieves near-clinical accuracy across six anatomical sites using a single, unified model. This method overcomes the computational bottlenecks and non-differentiability of traditional algorithms. Beyond serving as a computationally consistent tool for independent verification, the framework could provide a differentiable engine essential for end-to-end deep learning-based automatic planning. With future extensions to additional beam energies, this model establishes a scalable foundation for next-generation open-source planning and adaptive delivery systems."